\documentclass[aps,pra,showpacs]{revtex4}
\usepackage{fixmath}
\usepackage{amsfonts}
\usepackage{graphicx}
\usepackage{amsmath}
\usepackage{setspace}
\usepackage{subfigure}
\def\>{\rangle}

\def\<{\langle}
\def\>{\rangle}

\begin{document}
\title{Unconditional Security of Single-Photon Differential Phase Shift Quantum Key Distribution}
\author{Kai Wen$^1$, Kiyoshi Tamaki$^{2,3}$, and Yoshihisa Yamamoto$^{1,4}$}
\affiliation{$^1$ Edward L. Ginzton Laboratory, Stanford University, Stanford, California 94305, USA\\
$^2$ NTT Basic Research Laboratories, NTT Corporation, 3-1
Morinosato Wakamiya Atsugi-Shi, Kanagawa, 243-0198, Japan\\
$^3$ CREST, JST Agency, 4-1-8 Honcho, Kawaguchi, Saitana, 332-0012,
Japan\\
$^4$ National Institute of Informatics, 2-1-2 Hitotsubashi,
Chiyoda-ku, Tokyo, 101-843, Japan}
\date{\today}
\begin{abstract}
In this Letter, we prove the unconditional security of single-photon
differential phase shift quantum key distribution (DPS-QKD)
protocol, based on the conversion to an equivalent
entanglement-based protocol. We estimate the upper bound of the
phase error rate from the bit error rate, and show that DPS-QKD can
generate unconditionally secure key when the bit error rate is not
greater than 4.12\%. This proof is the first step to the
unconditional security proof of coherent state DPS-QKD.
\end{abstract}
 \maketitle

Quantum key distribution (QKD) protocols are one of the most
important applications of quantum information theory. Great efforts
have been devoted to prove the unconditional security of these
protocols through noisy channels. The first QKD protocol, named
Bennett and Brassard 1984 (BB84) protocol\cite{bb84}, has been
proven to be unconditionally secure\cite{bb84_sec, shor}.

In 2002, Inoue et al. proposed the differential phase shift quantum
key distribution (DPS-QKD) protocol\cite{Inoue}, where Alice encodes
the key bits by preparing the relative phase shifts between two
consecutive pulses in $0$ or $\pi$ and Bob employs a one-bit delay
Mach-Zehnder (M-Z) interferometer to retrieve the key from the phase
shifts. The advantages of DPS-QKD mainly lie in its simple and
robust experimental implementation. Only one measurement basis is
involved in the protocol, and thus the experiment requires minimum
setup, namely, one source and two detectors. Also, for the BB84
protocol, Alice should generate a random base string to encode the
key and Bob also needs to randomly select his measurement bases.
However, in DPS-QKD, they do not need to perform these two steps.
Furthermore, DPS-QKD utilizes the relative phases of the pulses
which are not affected by the birefringence in fibers. Finally, by
using coherent sources, DPS-QKD is secure against
photon-number-splitting attack, because they can be
detected\cite{pns_dps}, while the BB84 protocol with coherent
sources requires intensity modulators to generate decoy states to
prevent such attack\cite{decoy}. Thanks to these simplicities,
experiments over long distances and with high bit rates have been
performed\cite{sspd, megabits}. On the other hand, whether DPS-QKD
is unconditionally secure is an important problem both from the
practical and theoretical aspects. So far, the security against
limited attacks such as the general attack for individual
photons\cite{ew} and the so-called sequential
attack\cite{seq_attack} have been analyzed.

In this Letter, as the first important step towards the security
proof of coherent state DPS-QKD, we present the proof of the
unconditional security of DPS-QKD with a single-photon source
against the most general attacks. This proof gives insight to the
underlying security properties of DPS-QKD. By analyzing an
equivalent entanglement-based DPS-QKD, we find that the phase error
rate of each time slot can be upper-bounded by the bit error rate of
the same time slot and its adjacent time slots. Thus, the
unconditional security is achieved by performing privacy
amplification based on the upper bound of the phase error rate.
Thanks to the equivalence, we can apply the results to the
prepare-and-measure DPS-QKD.

Before constructing the entanglement-based protocol, we define the
encoded states in the prepare-and-measure DPS-QKD. With the
single-photon source, Alice splits the single-photon wavepacket into
$n$ pulses with identical amplitudes to form a block. Particularly,
the state before encoding the secret key is
$|\phi_0\>=\frac{1}{\sqrt{n}}\sum_{k=1}^{n} a_k^\dag |\textrm{vac}\>
=\frac{1}{\sqrt{n}}\sum_{k=1}^{n} |D_k\>$, where $a_k^\dag$ is the
creation operator of the pulse in the $k$-th slot and
$|D_k\>=a_k^\dag|\textrm{vac}\>$. Then following the proposal of
DPS-QKD, Alice encodes an $(n-1)$-bit random secret key into this
block. For an $(n-1)$-bit random but fixed integer $j$, we express
its $(n-1)$-bit binary format as $( j_1 j_2 \cdots j_{n-2} j_{n-1}
)_2$. Then the encoded state of the block of a single photon is,
$|\phi_{j}\>=\frac{1}{\sqrt{n}}\left[|D_1\>+\sum_{k=2}^{n}
(-1)^{j'_{k-1}} |D_k\>\right]$, where $j'_k = \sum_{l=1}^k j_l$.

Given the above encoding scheme, we can construct the corresponding
states in the entanglement-based protocol. The equivalence between
the entanglement-based and the prepare-and-measure protocols is
obtained by following the technique by Shor and Preskill\cite{shor}.
For each encoding block Alice prepares additional $(n-1)$ qubits
which are stored without disturbances in her own quantum memory
throughout the protocol. These qubits, labeled with $A_1, \cdots,
A_{n-1}$, are entangled with a single photon, labeled $B$ in the
corresponding block, which are described as
\begin{equation}\label{initial_edps}
|\phi\>= \frac{1}{\sqrt{2^{n-1}}} \sum_{j=0}^{2^{n-1}}\left[
(|j_1\>_{A_1} \cdots |j_{n-1}\>_{A_{n-1}}) \otimes |\phi_{j}\>_B
\right].
\end{equation}

On Bob's side, after he receives the single photons, he first
applies quantum non-demolition (QND) measurement to determine the
number of incoming photons in a block and discard the blocks with
multi photons or the vacuum. This step is necessary because the
following entanglement purification protocol requires a well-defined
qubit on his side. The QND measurement commutes with all other
operations on Bob's side. Therefore, in the prepare-and-measure
protocol, Bob can replace the QND measurement by the photon number
resolving (PNR) detectors which are capable of discriminating the
vacuum, the single-photon state, and multi-photon states.
Particularly, by using two PNR detectors, Bob only accepts the
instances when one detector obtains the single-photon state and the
other one obtains the vacuum state in one block.

\begin{figure}
\begin{center}
\includegraphics[width=8cm]{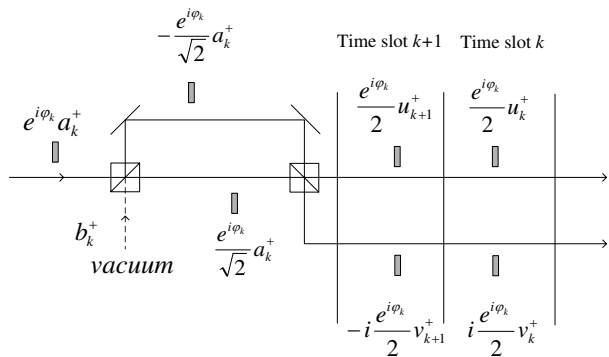}
\caption{Schematics of the 1-bit delay M-Z
interferometer}\label{fig_DPS}
\end{center}
\end{figure}

Then, the single photon goes to a 1-bit delay M-Z interferometer,
shown in Fig. \ref{fig_DPS}. An incoming photon state in $|D_k\>$,
is split into four different photon states in the two output ports
and two consecutive time slots. Particularly, we obtain $a_k^\dag
\Rightarrow \frac{1}{2}(u_k^\dag + i v_k^\dag + u_{k+1}^\dag - i
v_{k+1}^\dag)$, where $u_k^\dag$ and $v_k^\dag$ are the creation
operators of the pulses of the time slot $k$ in the two output
ports. For convenience, Bob applies a $\pi/2$ phase rotation on each
pulse in the bottom output port. Given $|U_k\> =
u^\dag_k|\textrm{vac}\>$ and $|V_k\> = v^\dag_k|\textrm{vac}\>$, the
operation of the interferometer ($M_{DPS}$) on each $|D_k\>$ can be
written as
\begin{eqnarray}
M_{DPS} |D_k\> = \frac{1}{2}(|U_k\> - |V_k\> + |U_{k+1}\> +
|V_{k+1}\>).
\end{eqnarray}

Bob further applies a hypothetical filtering operation in order to
project the single photon into a two-level state required for the
entanglement purification protocol. The filter operation is
described by a set of Kraus operators $F = \{F_1, F_2, \cdots,
F_n\}$, namely $F_l = |U_l\>\<U_l|+|V_l\>\<V_l|, \textrm{for}\
l=2,\cdots,n$ and $F_1 = I - \sum_{l=2}^{n} F_l$ in which $I$ is the
identity matrix, namely,
$I=\sum_{l=1}^{n+1}(|U_l\>\<U_l|+|V_l\>\<V_l|)$. Note that the
projection operators commute with each other and represent
monitoring the time slots of the detection events. Bob then publicly
announces which time slot the photon was projected to. Alice and Bob
will discard the inconclusive blocks where Bob obtains $F_1$. By its
projection to a certain time slot $l$, the photon is in a
well-defined qubit state. Particularly, we can define the $Z$-basis
of the photon as $\{|U_l\>, |V_l\>\}$, representing whether it
travels along the top or bottom output port of the interferometer.
Accordingly, we define Bob's Pauli operators as $Z_{B_l} =
|U_l\>\<U_l|-|V_l\>\<V_l|$ and $X_{B_l} =
|U_l\>\<V_l|+|V_l\>\<U_l|$.

Finally, Alice and Bob should tackle the eavesdropping and the
channel errors. On the one hand, when the channel is ideal and no
eavesdropping exists, it is easy to show that Alice and Bob obtain a
maximally entangled pair from each projected qubit. Particularly,
Alice discards all the qubits on her side with the label other than
$l-1$, in which $l$ is Bob's projection outcome. Mathematically, it
is equivalent to partially tracing these qubits in the state of Eq.
(\ref{initial_edps}). Combining with Bob's filter projection, Alice
and Bob share the Bell state, namely, $
|\Phi^+\>=\frac{1}{\sqrt{2}}(|0\>_{A_{l-1}} \otimes |U_l\>_B +
|1\>_{A_{l-1}} \otimes |V_l\>_B)$.

On the other hand, if the channel is noisy or there is an
eavesdropper, Alice and Bob share a corrupted two-qubit state. In
this case, Alice and Bob employ an appropriate entanglement
purification protocol based on Calderbank-Shor-Steane (CSS)
code\cite{css}, to distill the Bell state. If the entanglement
purification protocol succeeds, the resulting smaller set of states
shared by Alice and Bob will have very high fidelity. Using the
argument that high fidelity implies low entropy\cite{LoChau} or
composability argument\cite{composability}, Alice and Bob can
generate an unconditionally secure key by measuring the distilled
states in their own respective $Z$-basis. Therefore, the key to the
unconditional security proof is whether they can estimate the bit
error rate and the phase error rate, which is necessary for choosing
an appropriate CSS code for the successful purification. As for the
bit errors, Alice and Bob can estimate them by using test bits.
However, in the prepare-and-measure protocol, since they cannot
directly measure the phase errors of the test bits, they have to
upper-bound them only from the observed quantities.

In what follows, we concentrate only on the untested bits, and for
the estimation of the phase error rate, we appeal to Azuma's
inequality\cite{azuma1,azuma2}. First, we define $p_{b,l}^{(k)}$ as
the probability of observing a bit error in the $l$-th time slot of
the $k$-th photon pair. We allow $p_{b,l}^{(k)}$ to be dependent on
the previous $k-1$ events, in other words, this probability is a
conditional probability. Moreover, we define $Ne_{b,l}$ as the
number of the actual bit errors in the $l$-th time slot after
N-photon-pair emission. Similarly, we can define the sequence
$p^{(k)}_{p,l}$ and $e_{p,l}$ for the phase errors in the $l$-th
time slot. A consequence of Azuma's inequality states that
$\textrm{Pr}\left[\left|e_{\Lambda,l} - \frac{\sum_{k=1}^N
p^{(k)}_{\Lambda,l}}{N}\right|\geq \epsilon \right] \leq 2
e^{-N\epsilon^2/2}$, for arbitrary positive number $\epsilon$, both
$\Lambda \in \{b,p\}$, and all conclusive time slots
$l$\cite{azuma2}. Therefore, if we can find the relation, $
\sum_{l=2}^{n} p^{(k)}_{p,l} \leq \sum_{l=2}^{n} C_l p^{(k)}_{b,l}$
for certain $C_2, \cdots, C_n$, the total phase error rate $e_{p} =
\sum_{l=2}^{n} e_{p,l}$ can be bounded by the same relation, namely,
$e_{p} \leq \sum_{l=2}^{n} C_l e_{b,l}$. On the other hand, the
random sampling theory states that $e_{b,l}$ is close to the
measured bit error rate on the test bits with an exponentially small
probability\cite{LoChau}.

Eve's most general attacks entangle the whole blocks with her
ancila. Focusing on one certain block, e.g., the $k$-th block, the
attacks can be reduced to a Kraus operator acting only on this
block, by the following two steps: Firstly, since Azuma's inequality
requires conditional probabilities, we suppose that Alice and Bob
have performed fictitious Bell measurements to test errors on the
previous $(k-1)$ blocks. We project all the previous systems
according to the outcomes and trace out the $(k-1)$ blocks.
Secondly, we trace out the $(k+1)$,-th $\cdots$, $n$-th blocks and
Eve's ancila. The resulting operator is $\Phi_k(\rho_k) = \sum_s
E_s^{(k)\dag} \rho_k E_s^{(k)}$, where $\rho_k$ is the unperturbed
state of the $k$-th block. Because the actual measurement outcomes
and Eve's coherent attacks are unknown, the operator $E_{s}^{(k)}$
is arbitrary and dependent on arbitrary measurement outcomes of the
previous $(k-1)$ blocks\cite{azuma2}.

The linearity of the Kraus operator allow us to consider only one of
its components, $E^{(k)}_s = (a_{ij})$, an arbitrary $(n \times
n)$-dimension matrix acting on the single-photon state. The
corrupted state of the $k$-th block becomes $E^{(k)}_s |\phi\>$. The
final state after Bob's interferometer and the filter operation is $
|\phi^{(k)}_l\> = F_l M_{DPS} E^{(k)}_s |\phi\>$. Therefore, for
this time slot, we can obtain the possibilities $p^{(k)}_{b,l} =
\<\phi^{(k)}_l| \frac{\mathbold{1} - Z_{A_{l-1}} Z_{B_l}}{2}
|\phi^{(k)}_l\>$ and $p^{(k)}_{p,l} = \<\phi^{(k)}_l|
\frac{\mathbold{1} - X_{A_{l-1}} X_{B_l}}{2} |\phi^{(k)}_l\>$, which
are conditioned on arbitrary previous events. The calculations show
that
\begin{eqnarray}
p^{(k)}_{b,l} &=\frac{1}{4n} [&|a_{l-1,l-1}-a_{l,l}|^2+
|a_{l-1,l}-a_{l,l-1}|^2 + (|a_{l-1,1}|^2+\cdots+|a_{l-1,l-2}|^2+
|a_{l-1,l+1}|^2+\cdots+|a_{l-1,n}|^2) +\nonumber \\
&& (|a_{l,1}|^2+\cdots+|a_{l,l-2}|^2 + |a_{l,l+1}|^2+\cdots+|a_{l,n}|^2)]\label{biterror},\\
p^{(k)}_{p,l} &= \frac{1}{2n} [&|a_{l,1}|^2+ \cdots + |a_{l,l-1}|^2
+ |a_{l-1,l}|^2+ \cdots + |a_{l-1,n}|^2]\label{pherror}.
\end{eqnarray}

By observing that for any two complex numbers $a$ and $b$,
$|a|^2+|b|^2 \leq \frac{3+\sqrt{5}}{2} (|a-b|^2+ |a|^2),$ we derive
that $2n\sum_{l=2}^{n}p^{(k)}_{p,l} - 4n \sum_{l=2}^{N}p^{(k)}_{p,l}
\leq [(|a_{1,2}|^2+|a_{2,1}|^2) - (|a_{1,2}-a_{2,1}|^2 +
|a_{2,1}|^2)]+ [(|a_{n-1,n}|^2+|a_{n,n-1}|^2) -
(|a_{n-1,n}-a_{n,n-1}|^2 + |a_{n-1,n}|^2] \leq
(1+\sqrt{5})/2[(|a_{1,2}-a_{2,1}|^2 + |a_{2,1}|^2)+
(|a_{n-1,n}-a_{n,n-1}|^2 + |a_{n-1,n}|^2)] \leq (1+\sqrt{5})/2
\times 4n\sum_{l=2}^{N}p^{(k)}_{b,l}$, or equivalently
$\sum_{l=2}^{n}p^{(k)}_{p,l} \leq (3+\sqrt{5})
\sum_{l=2}^{n}p^{(k)}_{b,l}$. It should be emphasized that the
relations are general for arbitrary matrix component and arbitrary
previous measurement outcomes. Using Azuma's inequality, we
therefore obtain
\begin{eqnarray}\label{phase_bound}
e_p \leq (3+\sqrt{5}) \sum_{l=2}^{n}e_{b,l} = (3+\sqrt{5})e_b,
\end{eqnarray}
where $e_b = \sum_{l=2}^{n}e_{b,l}$ is the total bit error rate over
all conclusive time slots. The derivation clearly demonstrates the
essence of DPS-QKD in which the upper bound of the phase error rate
in certain time slot can only be estimated by combining the bit
error rates in the same and adjacent time slots. Note that this
upper bound does not apply to the $n=2$ case in which the phase
error rate can be as high as 50\% and we have no chance to generate
the secret key. This is the case where Alice uses two orthogonal
states and Eve has free access to the information. The upper bound
is valid for $n \geq 3$ in which the states are mutually
non-orthogonal and no unambiguous state discrimination exists. This
results share some similarity in the security proof of the Bennet
1992 protocol\cite{b92}.

Combining the above three arguments, we can derive the
unconditionally secure key rate of the entanglement-based DPS-QKD
and the single-photon DPS-QKD, namely,
\begin{eqnarray}
R_{DPS} \geq p_{DPS}\left[1 - H(e_b) - H((3+\sqrt{5})e_b)\right],
\end{eqnarray}
where $H(x)$ is the binary Shannon entropy, namely, $H(x)=-x
\log_2(x)-(1-x) \log_2(1-x)$, and $p_{DPS}$ is the conclusive
detector click rate per pulse in DPS-QKD.

Finally, we compare the key rates of the unconditionally secure BB84
protocol ($R_{BB84}$)\cite{shor}, DPS-QKD against general attack for
individual photons ($R_{IND}$)\cite{ew}, and unconditionally secure
DPS-QKD ($R_{DPS}$). We assume single-photon sources in all three
protocols. In the presence of channel losses, we express that
$R_{BB84} = p_{BB84}\left[1 - 2 H(e_b)\right]$, and $R_{IND} =
p_{IND}\left\{-\log_2\left[1 - e_b^2 - \frac{(1-6 e_b)^2}{2}\right]
- H(e_b)\right\}$, where $p_{BB84}$ and $p_{IND}$ are the conclusive
detector click rates per pulse in the corresponding protocols. Here
we adopt the result of DPS-QKD against general attack for individual
photons\cite{ew} to the case with a single-photon source. We assume
that the coding efficiency for the bit error correction approaches
to Shannon limit in all three cases. When $R_{BB84}$, $R_{IND}$ and
$R_{DPS}$ hit zero, the upper bound of the tolerable bit error rates
for three protocols are found to be $11\%$, $6.09\%$, and $4.12\%$
respectively. Note that in DPS-QKD the phase error rate is
indirectly estimated by $M_{DPS}$ and the filter while it is
directly estimated in the BB84 protocol. This is an essential
insight we obtained in this Letter, and this poor estimation results
in lower error rate threshold of DPS-QKD compared to the one of the
BB84 protocol.

To simulate the resulting key rates, we take the parameters from
Ref. \cite{sspd}, where the dark count rate, the time window and the
baseline error rate are 50 Hz, 50 ps and 2.3 \% respectively.
Therefore, the dark count rate per detector per time slot is $d =
2.5 \times 10^{-9}$. We assume that all protocols use two detectors.
So $p_{BB84} = (\eta + 2 d)/2$, where $\eta$ is the total efficiency
including the channel, the detectors and all other devices. In
DPS-QKD, we further assume that the loss event happens equally on
every pulse in the transmission, so that the probability of getting
a conclusive event is $(n-1)/n$. On the other hand, a dark count can
occur in every time slot with equal probability. Therefore, by
noting that Bob obtains at most 1 photon out of a block with $n$
pulses, $p_{IND} = p_{DPS} = \eta(n-1)/n^2 + 2d(n-1)/n$. Moreover,
the bit error rates can be modeled as $e_b = (e \eta + d)/(\eta +
2d)$ for the BB84 protocol and $e_b = [e \eta(n-1)/n^2 +
d(n-1)/n]/[\eta(n-1)/n^2 + 2d(n-1)/n]$ for DPS-QKD, where $e$ is the
baseline error rate given above.
\begin{figure}
\begin{center}
\subfigure[] { \label{fig_single_result}
\includegraphics[width=8cm]{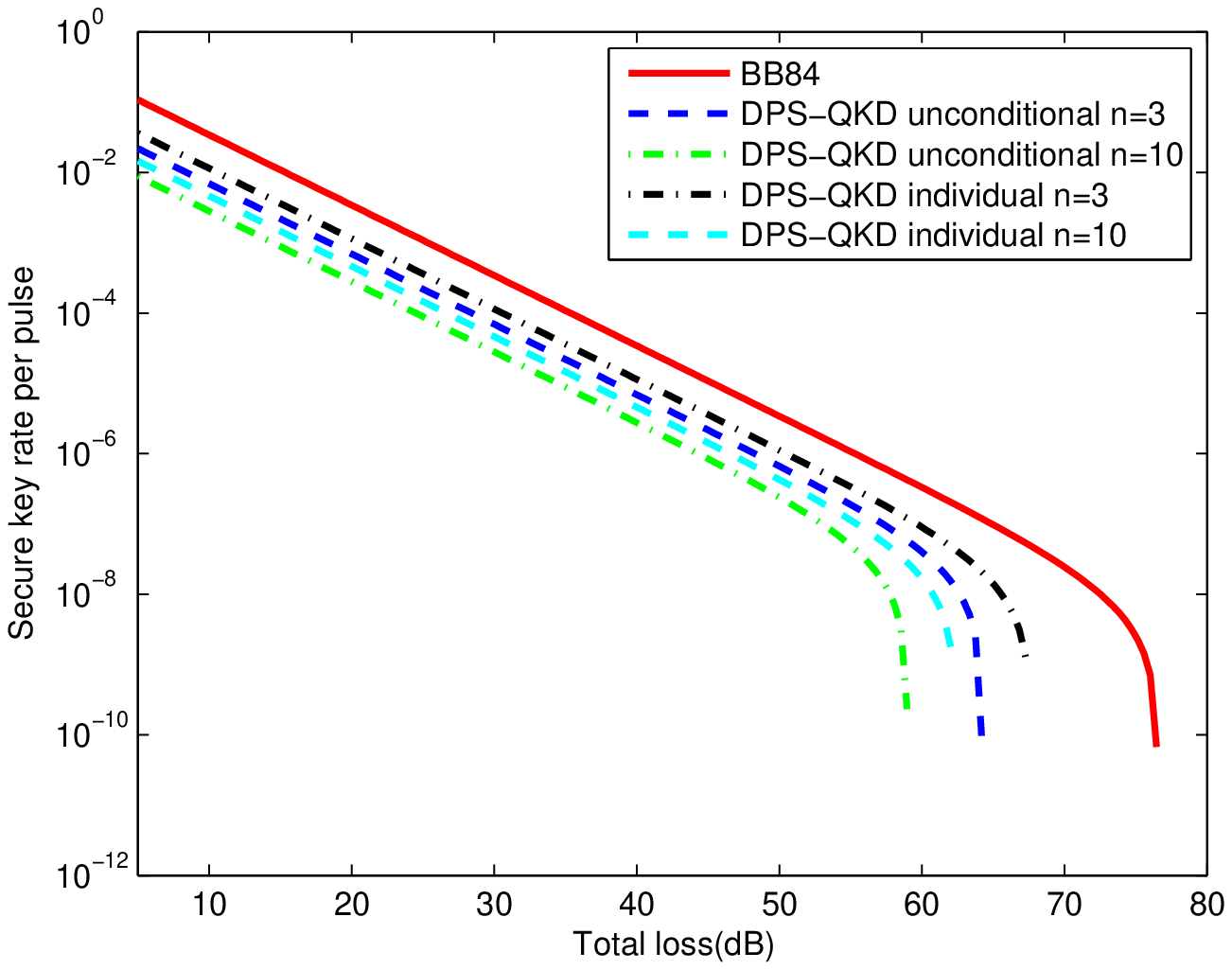}
}
\subfigure[] { \label{fig_ee}
\includegraphics[width=8cm]{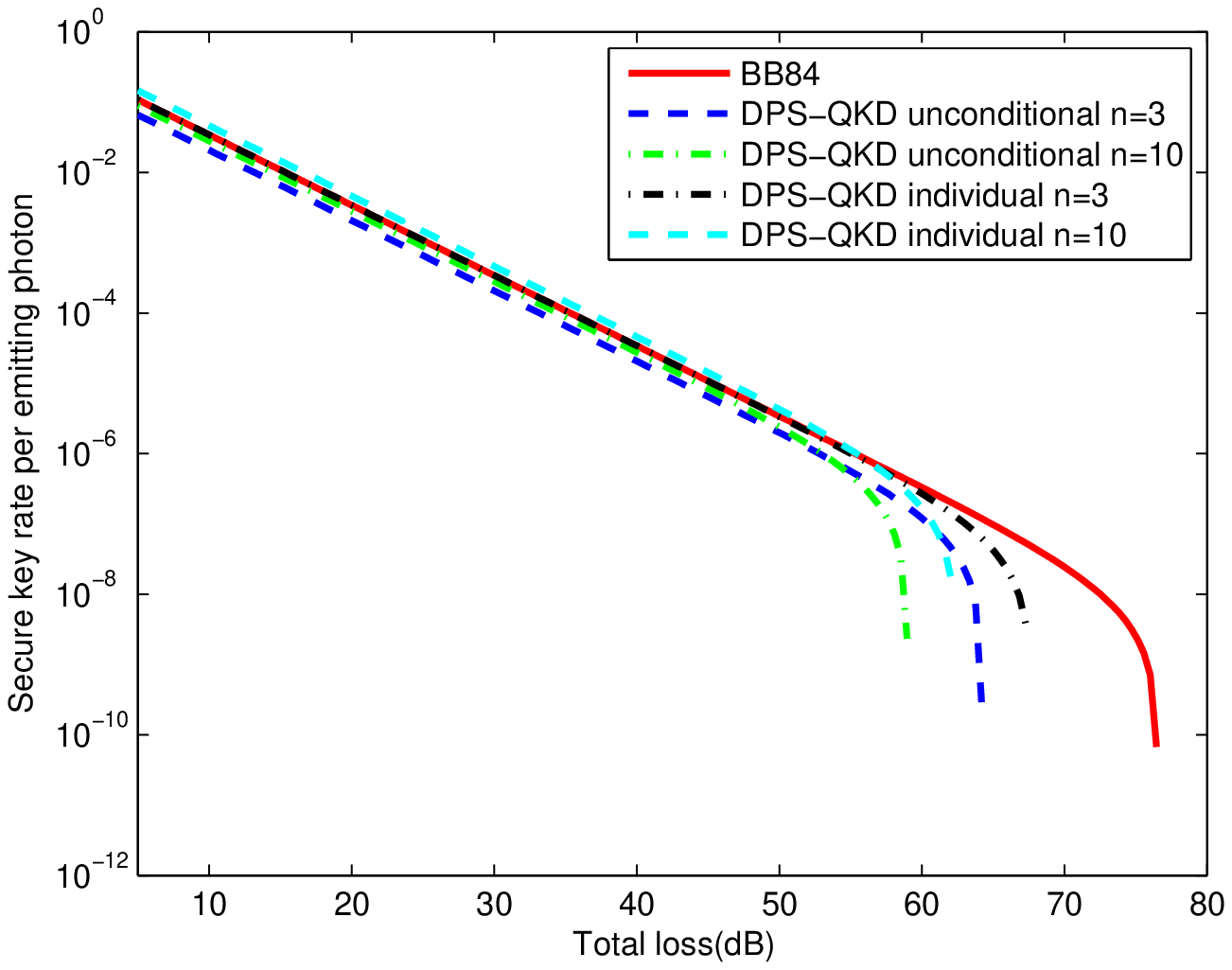}
} \caption{Secure key rates per pulse \subref{fig_single_result} and
per emitting photon \subref{fig_ee} as a function of the total loss.
Red solid line: $R_{BB84}$; blue dashed line and green dash-dot
line: $R_{DPS}$ with $n=3, 10$ respectively; black dash-dot line and
light blue dashed line: $R_{IND}$ with $n=3, 10$ respectively.}
\end{center}
\end{figure}

Fig. \ref{fig_single_result} and \ref{fig_ee} illustrate the secure
key rates per pulse and the energy efficiencies, namely, the secure
key rates per emitting photon. As expected, the unconditionally
secure key rate and the upper bound of the tolerable bit error rate
of DPS-QKD are lower than those of DPS-QKD against general attack
for individual photons. From Fig. \ref{fig_ee}, larger $n$ has
higher energy efficiency because every photon received by Bob has
lower chance to be discarded. However, larger $n$ will decrease the
probability of getting a signal from a pulse and increase the dark
count rate per block, and thus lead to lower secure key rate per
pulse and lower achievable distance as shown in Fig.
\ref{fig_single_result}. Based on these observations, we find that
$n=3$ yields the optimal secure key rate per pulse and the maximum
achievable distance.

In conclusion, we have proven the unconditional security of DPS-QKD
with a single-photon source and evaluated its secure key rate. The
security is based on the non-orthogonality of the encoding states
for $n \geq 3$ and Bob's 1-bit delay operation. We hope that our
unconditional security proof is a first step toward the security
proof of coherent state DPS-QKD.

The authors wish to thank Norbert L\"{u}tkenhaus, Masato Koashi,
Daniel Gottesman, Hoi-Kwong Lo, Qiang Zhang, and Hiroki Takesue for
very fruitful discussion on the topic of this Letter. This research
was supported by NICT, the MURI Center for Photonic Quantum
Information Systems (ARMY, DAAD19-03-1-0199), NTT Basic Research
Laboratories, SORST, CREST programs, Science and Technology Agency
of Japan (JST), and Hamamatsu Photonics.

\end{document}